\documentclass[11pt,a4paper]{article}

\renewcommand{\author}{Emil Khalisi}
\newcommand{\titel}{The Double Eclipse at the Downfall of Old Babylon}
\newcommand{\version}{Version 1.54}
\renewcommand{\date}{\today}

\usepackage{hyperref}
\hypersetup{pdfauthor={Emil Khalisi}}
\hypersetup{pdftitle={\titel}}
\hypersetup{pdfsubject={Online publication, \version\ of \date }}
\hypersetup{pdfkeywords={Eclipses, Astronomical Dating, Chronology,
    Babylon, Akkad, Ur, Gutian, Venus tablet, EAE 20-XI}}

\usepackage[T1]{fontenc}        
\usepackage{mathptmx}           
\usepackage{pifont}             
\usepackage[scaled=0.92]{helvet} 
\usepackage[british]{babel}     
\usepackage{amsmath}            
\usepackage{microtype}          

\usepackage{hhline}             
\usepackage{arydshln}           
\usepackage{nameref}            

\usepackage{calc}
\usepackage[a4paper]{geometry}  
\usepackage{scrextend}           
\changefontsizes{10pt}
\usepackage{titlesec}
\titleformat*{\section}{\large\bfseries}
\titleformat*{\subsection}{\normalsize\bfseries}

\usepackage{graphicx}           
\usepackage{float}              
\usepackage{caption}            
\usepackage{fancyhdr}
\renewcommand{\headrulewidth}{0.4pt}
\usepackage{colortbl}             
\definecolor{grey20}{RGB}{208,208,208}
\definecolor{grey10}{RGB}{230,230,230}
\definecolor{braun}{RGB}{230,141,000}
\definecolor{blaua}{RGB}{186,217,255}
\definecolor{blaub}{RGB}{152,198,255}
\definecolor{tuerkis}{RGB}{118,228,255}
\definecolor{hellblau}{RGB}{198,255,252}
\definecolor{hellgelb}{RGB}{255,247,123}

\newcommand{\guti}{\cellcolor{braun}}
\newcommand{\urukiv}{\cellcolor{blaua}}
\newcommand{\urukv}{\cellcolor{blaub}}
\newcommand{\urdrei}{\cellcolor{tuerkis}}
\newcommand{\urnammu}{\cellcolor{hellblau}}

\begin{document}


\fancyhead{}
\fancyhead[LO]{%
   \footnotesize \textsc{In original form published in:} \\
         Habilitation at the University of Heidelberg
}
\fancyhead[RO]{
   \footnotesize {\tt arXiv: 0000.00000 [physics.hist-ph]}\\
   \footnotesize {Date: 14th July 2020}%
}
\fancyfoot[C]{\thepage}

\renewcommand{\abstractname}{}

\twocolumn[
\begin{@twocolumnfalse}

\section*{\centerline{\LARGE \titel }}

\vspace{\baselineskip}
\begin{center}
{\author \\}
\textit{69126 Heidelberg, Germany}\\
\textit{e-mail:} \texttt{ekhalisi@khalisi.com}\\
\end{center}

\vspace{-\baselineskip}
\begin{abstract}
\changefontsizes{10pt}

\noindent
\textbf{Abstract.}
For many decades scholars converse how to correctly include the Old
Babylonian Empire into the absolute timeline of history.
A cuneiform text from the series \textit{Enuma Anu Enlil} (EAE \#20)
reports on the destruction of Babylon after a lunar and solar
eclipse.
Eclipses provide a great tool for examining historic events,
and this account will be our basis for the investigation of eclipse
pairs to be fitted into the various chronologies proposed.
We consider three interpretations of that text:
literal understanding, the inverted sequence of the two eclipse
types, and a relocation of the setting to Akkad.
All variants show imperfections.
The least complications emerge when the account draws upon the
third option, i.e.\ if it is allotted to 2161 BCE, as the year would
mark the end of the Gutian rule in Akkad.
But in any case the account seems to support rather the Long
Chronology than any other.
When dealing with eclipses such far back in time, we also allow
for the shift of the visibility zone in regard to the irregular
deceleration of Earth's rotation.\\

\noindent
\textbf{Keywords:}
Astronomical dating, Chronology, Solar eclipse, Earth's rotation,
   Babylon, Akkad.


\end{abstract}

\vspace{\baselineskip}
\centerline{\rule{0.8\textwidth}{0.4pt}}
\vspace{2\baselineskip}

\end{@twocolumnfalse}
]



\section{Introduction}

The Babylonian civilisation stands at the origin of scholarship
that has left traces right to the present.
As the peoples of Mesopotamia laid the foundation for our modern
science, its culture emanated to neighbours as well as successors.
The history of Babylon can be divided into two main phases:
the Old Babylonian period at the beginning of the second millennium
BCE and the Neo-Babylonian from 626 to 539 BCE.
The dating of the Old Babylonian Empire causes much trouble.
Depending on the field of study, its absolute time varies within
a range of two centuries.

In this paper, we briefly delineate the political milestones of
Mesopotamian history and explore the timescales in use that rest
upon different methods of measurement.
These timescales are the key element when taking a closer look at
the history of the ancient world.
We will not provide another new chronology, but try to fit an
account on an eclipse pair into the existing ones.
This account describes the downfall of Babylon after a lunar and
a solar eclipse occurring in the same month.
Eclipses often recur pairwise, but for the same location they are
rare.
As our results fall into various eras, we ponder over the most
likely chronology.
Finally, we will end up in Akkad as the probable solution and make
a suggestion for a slight restructure in the scheme of rulers
there.

Moreover, when analysing historical eclipses, we do consider the
irregular fluctuations in the earth's rotation giving rise to
an error decisive for the eclipse to be visible in a certain
region.
Though a reliable answer is blurred by the error, our findings
will not alter the historical framework.


\section{Overview of the Mesopotamian History}

The first forms of a broader government along the two rivers
Euphrates and Tigris go as far back as the third millennium BCE
when various city states rose with local dynasties.
They were frequently at odds with each other, as inscriptions on
steles and temple fragments prove.
A territorial state probably formed in the second half of the 3rd
millennium.
The history of Mesopotamia commences with the legendary king Sargon
of Akkad who conquered a considerable realm in that region.
Until now the geographic location of Akkad is not well known,
probably it is to be placed in the vicinity of today's Baghdad.
Other dynasties existed somewhat farther to the south in Uruk, Ur,
Kish, and Lagash.

Sargon is regarded as the first notable monarch in Mesopotamia.
He paved the way for the later empires of Babylon and Assyria.
Prior to him there were godlike kings reigning for many thousands
of years, some 30,000 or even 200,000 years.
Their lifetimes cannot be determined, in general.
One of them is Gilgamesh, the king of Uruk, who is linked to the
legend of the Noachian Flood.
That old time before Babylon is often bundled to the
``Sumerian Era''.

\enlargethispage{\baselineskip}

The Sumerians laid the scientific basis of the advanced civilisation
that propagated throughout the entire region.
Mathematics was based on a sexagesimal system (our clocks and angles
still maintain that custom);
writing was performed on clay tablets;
astronomy compiled and named star constellations;
and calendrical adjustments of lunar months were practised within
the cycle of seasons.
Venus was adored as the star of the goddess Inanna
(\glqq Mistress of Heavens\grqq )
and called ``Ishtar'' by the Akkadians.
She retained the role of being one of the most important heavenly
bodies much later when the power passed over to the Babylonians.

The foundation of Babylon presumably dates back to some time
during the Sargonic dynasty.
It existed as a small village at about 2000 BCE
till a late Sumerian king established the first dynasty there and
raised it to be his capital.
Thereupon it continued to grow to a cultural center.

The most prominent king of the first Babylonian dynasty was
Hammurabi.
He lived in the first half of the second millennium.
His regency lasted for 43 years, and he established the
significance of the then puny city state.
His antecessors expanded the region of influence, but Hammurabi
enforced domestic policies like building an infrastructure,
installing ambassadors with the neighbours, construction of
religious temples, and the famous code, an early collection of
legal rules valid for everyone.
By skilful tactical manoeuvres he conquered the former empires
of Akkad, Uruk, and Ashur during his reign.
Individual wealth increased, and the number of inhabitants in
Babylon, too.
Under his rule it became the dominant power in Mesopotamia.

The prosperity of Old Babylon lasted for $\approx$300 years till
it ceased after one dynasty only.
The dynasty incorporated 11 kings, of which Hammurabi was the
sixth.
After him the empire lost influence quite rapidly.
Ammi-saduqa was the fourth successor of Hammurabi, and thereafter
his son ascended the throne as 11th king before the kingdom
collapsed.
Babylon was captured by the Hittite king Mur\v{s}ili I.
Written documents suspended afterwards.

The Hittites were not able to sustain the city, because of its
large distance from their heartland, and abandoned it.
The Kassites, a tribe from the Zargos Mountains in the east,
occupied the power vacuum.
Changing sovereigns followed, some of them unknown, and probably
kingless years.
Nevertheless, the Kassites also contributed to the later culture
of the Neo-Babylonian Empire that was to flourish in the 7th
century BCE.
At about 1200 BCE, the Bronze Age ended with the collapse of the
Hittite Empire, and the Kassites were overthrown at about 1160
BCE as well.

%
\fancyhead{}
\fancyhead[CE, CO]{\footnotesize \itshape E.\ Khalisi (2020) : \titel}
\renewcommand{\headrulewidth}{0pt}


\section{The Main Chronologies}

When it comes to a more exact dating, any of the historical stages
above turn out uncertain on the absolute timescale.
It will be essential to reconcile them with other events, rulers,
and dynasties of neighbouring countries.
Taken the regencies alone, they prove inconsistent.
There exist dozens of lists winding up crosswise and imprecise,
patchy, and, most of all, contradictory.
Some texts incidentally contain an astronomical hint, like the
sighting of a planet or a moon phase, and one tries to conflate
that into a plot.

In the course of research different methods of reckoning have been
suggested, and the outcome led to a system referred to as ``choice
of chronology''.
The chronologies are an attempt to arrange some cornerstones of
the Babylonian history with the timeline of the neighbours as
the Egypts and Hittites.
They are no more than a time mesh, and their emergence is a complex
issue beyond the scope of this paper.
The origin is unfurled by Weir \cite{weir_1972} and Fotheringham
\cite{fotheringham_1928} in more detail.
Without discussing the concept, though virtually important, a few
basics are necessary to be summed up now.

\begin{table*}[t]
\caption{Various methods on dating the fall of the Babylon-I-Dynasty.
    The four main chronologies are highlighted.
    Years are given in BCE.
    For Ammi-saduqa's year 1 raise the year in the first column by 51.}
\label{tab:chronologies}
\centering
\begin{tabular}{clll}
\hline
\rowcolor{grey20}
 Year  & Evidence or method           & Author & see Ref. \\
\hline
%
%
 1926 & Venus tablet (EAE 63)         & Kugler (1912) & \cite{fotheringham_1928, weir_1972} \\
 1870 & Venus tablet (EAE 63)         & Fotheringham (1928) & \cite{fotheringham_1928, schoch_1925} \\
\cdashline{1-4}[0.5pt/5pt] 
%
 1733 & Assyrian king list            & Landsberger (1954) & \cite{eder_2004} \\
 1665 & Assyrian building inscriptions & Eder (2004) & \cite{eder_2004} \\
\cellcolor{grey10}
      & Statistics of astronomical data & Huber (1982/2000) & \cite{fru_huber_2000} \\
\cellcolor{grey10} \raisebox{0.75\height}[0mm][0mm]{1651}
      & Venus visibility + lunar months & Weir (1982) & \cite{weir_1982} \\
%
 1596 & ($\pm$7) Radiocarbon / K\"ultepe & Manning {\it etal} (2001) & \cite{fru_manning_2001} \\
\cellcolor{grey10}
      & Excavations in Alalakh/Turkey & Smith/Ungnad (1940) & \cite {weir_1972} \\
\cellcolor{grey10}
 1595 & Generation count of kings     & Rowton II (1958) & \cite{rowton_1958} \\
\cellcolor{grey10}
      & Political history             & MacQueen (1964) & \cite{macqueen_1964} \\
 1587 & Solar eclipse of Shamshi-Adad & De Jong (2013) & \cite{fru_dejong_2013} \\
 1587 & Volcano eruption on Thera/Santorini & De Jong (2010) & \cite{dejong-foertmeyer_2010} \\
 1560 & ($\pm$106) Radiocarbon / Nippur & Libby (1955) & \cite{rowton_1958, weir_1982} \\
 1557 & Solar eclipse on EAE \#24      & Henriksson (2005) & \cite{henriksson_2005} \\
 1547 & Ur III lunar eclipses + EAE 20 & Banjevic (2006) & \cite{banjevic_2006} \\
%
 1539 & Assyrian king list (Shamshi-Adad) & Boese (2012) & \cite{boese_2012} \\
\cellcolor{grey10}
      & Hittite documents             & Wilhelm \& Boese (1987) & \cite{wilhelm-boese_1987} \\
\cellcolor{grey10}
      & Average reign lengths of kings & Rowton I (1952) & \cite{rowton_1952} \\
\cellcolor{grey10} \raisebox{0.75\height}[0mm][0mm]{1531}
      & Venus tablet (EAE 63)          & van der Waerden (1968) & \cite{waerden_1968} \\
\cellcolor{grey10}
      & Traditional historical texts  & Cornelius (1942) & \cite{weir_1972} \\
 1523 & Venus tablet (EAE 63)          & Mebert (2010) & \cite{mebert-dejong} \\
%
\cellcolor{grey10}
      & Archaeological pottery        & Gasche {\it etal} (1998) & \cite{fru_gasche-etal_1998} \\
\cellcolor{grey10} \raisebox{0.75\height}[0mm][0mm]{1499}
      & Eponym lists from Mesop.\ + Egypt & Gertoux (2013) & \cite{fru_gertoux_2013} \\
  1467 & (suitablet chron.\ for EAE 63) & (statistics) & \cite{fru_huber_2000} \\
%
 1384 & ($\pm133$) Radiocarbon / Uruk  & M\"unnich (1957) & \cite{waerden_1968, weir_1982} \\
 1368 & Removal of ``shadow reigns''   & Furlong (2007) & \cite{furlong_2007} \\
 1362 & Venus (EAE 63) + month lengths & Mitchell (1990) & \cite{fru_mitchell} \\
\end{tabular}
\end{table*}

There are four variants for the chronology of the Near East:
long, middle, short, and ultra-short.
Each defines a few fixing points like the year of accession of
Hammurabi or the sack of Babylon.
None of the proposed chronologies is perfect.
The long one preferably agrees with astronomical data.
For historians it causes headaches, for they do not like it in
view of their king lists.
They favour the middle chronology, although it is subject to the
strongest criticism.
The short chronology adjusts to astronomy second best, but the
congruence of historical documents with the backreckoning turns
out of moderate use.
The fourth, ultra-short one came into being upon archaeological
studies of pottery and ceramics in 1998.
It does not fit to astronomy at all.
Taking the sack of Babylon as an example, the chronologies fix
this at the years 1651, 1595, 1531, or 1499 BCE, respectively,
see Table \ref{tab:chronologies}.
Apart from the main chronologies there are several other
suggestions of lesser significance.

The downfall of Babylon is equivalent with the dating of the
so-called ``Venus-Tablet'' (EAE \#63) from the era of the 10.\ king
Ammi-saduqa mentioned above.
Most scientists consider this tablet, that encloses the heliacal
risings and settings of Venus in a time interval of 21 years, as a
landmark for placing his lifetime in its proper historical context.
The Venus data must, however, get along with the lunar months,
that is to say, a reconstruction of the Babylonian calendar is
inevitable.
Each Babylonian month began at the visibility of the first crescent,
about 1 or 2 days after new moon, in the evening time, and it ended
with the observation of the next sickle after 29 or 30 days.
This puts certain constraints to the rising and setting times
of the moon.
From the astronomical point of view, the combination of the Venus
and Moon data will repeat every 8, 56, and 64 years, as such will
be the case after 112 and 120 years.

The periods should not be taken too simple, because the cycles are
incommensurable and are not related to each other.
The compatibility of the Venus data has to be determined
individually for each chronology of choice.
If the back-calculated data corresponds to the Venus tablet, it
is called a ``solution''.
It reduces the possible options for Ammi-saduqa's first year of
reign.
The ultra-short chronology, for example, does not provide a
solution to the Venus Tablet, but the requirements would fit
better 32 years later or earlier \cite{fru_huber_2000}.
On the whole, the end of the Babylon-I-Dynasty is placed 51 years
after Ammi-saduqa's accession.

The markers fixed by astronomy have to be balanced with further
aspects from history.
Some researchers make a detour via contemporary rulers of Hammurabi,
others rely on reports about long-lasting natural phenomena,
the next ones utilise economic texts,
while again others compute king lists extending over many centuries.
Attempts were also made with dendrochronology and the radiocarbon
method.
All interconnections turn out ambivalent, and sometimes they
confuse even more, as the affiliation of the sample is obscure.
Because of the rareness of evidence from the old Babylonian era,
any dating bears a lot of pitfalls, and the entire research
resembles a jigsaw puzzle.


\section{Eclipse Omen on EAE 20-XI}

An astrologic portent on the cuneiform tablet from the series
\textit{Enuma Anu Enlil} (EAE) is our starting point to determine
the destruction of Babylon.
The series of tablets contains hundreds of prophecies based on
weather phenomena, shape of clouds, halos, planets, day numbers,
and eclipses, of course.
For example, the dark side of the obscured disk (West, South, etc.)
was related to the direction what kingdom was affected by the
prophecy.
It would suffer from famine, rebellion, deluge, or the like.
The oldest EAE omens go back to the Akkadian and old Babylonian
times, most originate from the 7th century, while the youngest
ones are estimated at 195 BCE \cite{fru_dejong_2013}.
It is anything but obvious to which period of history the texts
apply.

\begin{figure*}[t]
\centering
\includegraphics[width=0.85\linewidth]{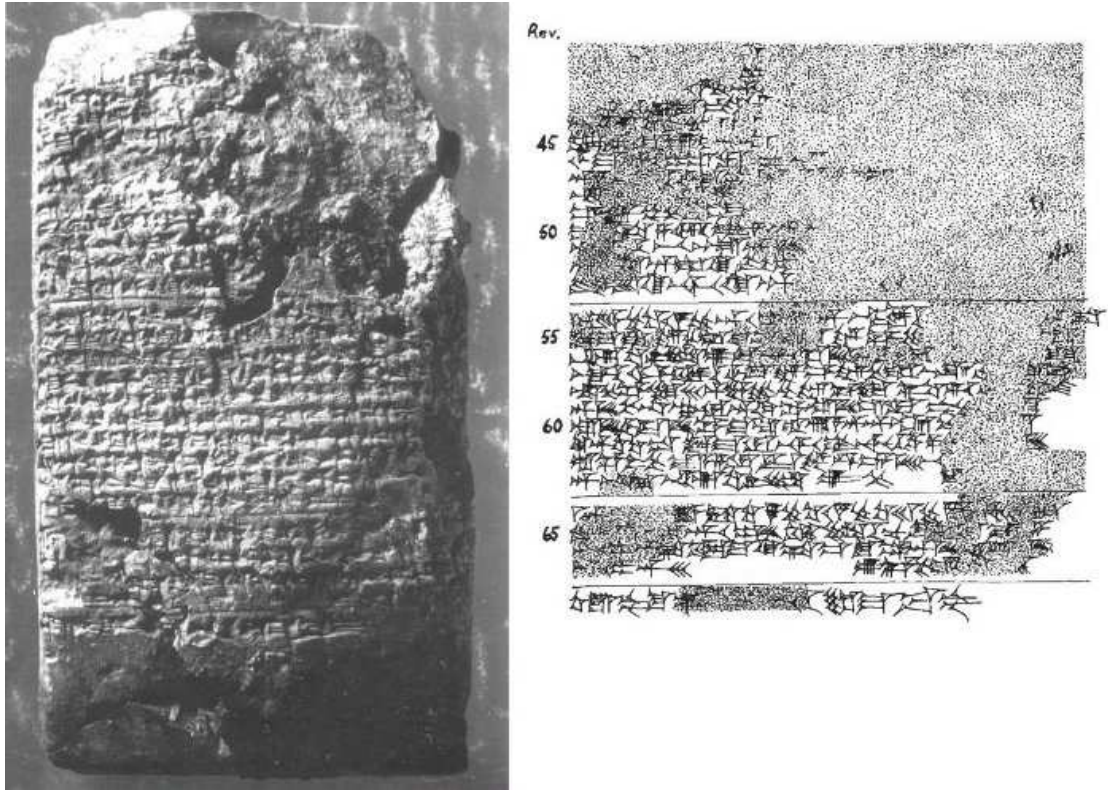}
\caption{Reverse of the EAE Tablet \#20 and the transcription of the
    symbols.
    Stock of the Iraq Museum, Baghdad, IM 124485 \cite{alrawi_2006}.}
\label{fig:eaetafel20}
\end{figure*}

Tablets \#15--22 are dedicated to lunar eclipses.
The content is composed of a protasis for the characteristics of
the eclipse plus an apodosis containing the portent.
The text is formulated as a conditional clause:
if the protasis occurs, then the apodosis will happen.
The compiler (or observer) did not aim at the investigation of
celestial interrelations but he wanted to find out what political
aftermath the eclipse would bring.
One can suppose that the portents ``stood the test of time'' by
having occurred once, at least, and, in a similar way, even more
than once.
Nevertheless, such loose information provides us with major
historic events that we would not be aware of otherwise.
In all, the records were edited and copied over and over again,
since they became part of an astrological lore.

The omens on our tablet of interest, EAE \#20, are arranged by
months (Fig.\ \ref{fig:eaetafel20}).
The lines 42 to 53 are damaged, but it was possible to reconstruct
the important segments by the aid of two other tablets.
There we read \cite{alrawi_2006, fru_huber_1999, fru_mitchell}:
%
\begin{quote}
If an eclipse occurs on the 14th day of Shabattu (month XI),
and the god, in his eclipse, becomes dark on the side south
above, and clears on the side west below;
the north wind [blows, and] in the dawn watch [the eclipse]
begins, and he (the moon) is seen with the sun.
His horns bend [toward] the sky.
His entire \textit{shurinnu} was not obscured, but disappeared.
On the 28th [day] you observe [the moon god] and an eclipse is
close by;
it begins and makes full [its time];
it (the \textit{shurinnu}) will show you the eclipse.
Observe his eclipse, [that of] the god who in his eclipse became
visible and disappeared, and bear in mind the north wind. ---
The prediction is given for Babylon:
the destruction of Babylon is near.
%
%
The king to whom Enlil said ``yes'', his people will be scattered.
His reign will end \dots\
[Ur] will take away from there [the hegemony of] Babylon.
Ur will take supremacy over Babylon.
If [night elapses] while the god is in eclipse:
[floods] will come [in the rivers], rains in the sky, the harvest
will be a success, good fortune will occur.
\end{quote}

In clearer words, the text deals with a pair of eclipses in the
time interval of 14 days.
On 14th and 28th day occurred a lunar, then a solar eclipse.
Shortly thereafter Babylon was destroyed, and the power passed
over to Ur.
The lunar eclipse began in the second part of the night
(``dawn watch'') and was in progress at sunrise.
The meaning of \textit{shurinnu} is not fully clear.
There are three possible explanations \cite{fru_huber_1999}:
as a partial eclipse, in general;
or that the moon set while partially darkened before it reached
totality;
or that it sank before being entirely restored to its full light.
Some consider \textit{shurinnu} as a technical term used in
Babylon for the ``crescent shaped appearance of an eclipsed moon''
\cite{alrawi_2006}.
Today we would just say ``partial''.

The ``14th day'' should not be interpreted too tight, because it
may stem from a systematisation of several events alike.
Depending on the start of the month, full moon can fall on the
13th or 15th day, too, and be eclipsed \cite{fru_huber_1999}.
Details about the solar eclipse two weeks later are absent.
The 28th day corresponds to the last visibility of the waning
crescent at dawn and would also be subject to systematisation.
Fortunately, the name of the month is given:
Shabattu is the 11th month of the Babylonian calendar equal to
our January or February.

An obstacle concerns the attribution of the 11th month.
The Babylonian calendar began in spring close to the vernal equinox
(month I).
Taking precession into account, it fell on 7th April in the middle
of the second millennium \cite{cornelius_1969}.
We do not know how good the old calendar was aligned to the
seasons.
Particularly, this concerns the month XII$_2$, the twelfth month
to be counted twice, to complete the season, and we can only
guess which years were affected by the insertion.
So, the new year started at some day between 20th March and 20th
April.
The author of this paper takes the liberty to render Shabattu
(XI) between the beginning of January and the end of March.


\section{Results for the Double Eclipse}

\begin{table*}[t]
\caption{Pairs of eclipses for Babylon (32.5$^{\circ}$ N, 44.4$^{\circ}$ E).
    Times are given in local time, LT = UT + 3h, corrected for the
    extrapolated $\Delta T$.}
\label{tab:babylonpaare}
%
\centering
\begin{tabular}{clrl|c|l}
\hline
\rowcolor{grey20}
\multicolumn{4}{c|}{\cellcolor{grey20}{Eclipse pair MS:}}& Chrono- &   \\
\rowcolor{grey20}
   & Date         &Time [LT]& Mag. & logy &
                              \raisebox{0.75\height}[0mm][0mm]{Remarks} \\
\hline
M: & 1753, Feb 28 &  1:07 & 0,378 &        &            \\
S: & 1753, Mrc 14 & 15:07 & 0,336 &        &
     \raisebox{0.75\height}[0mm][0mm]{too early?} \\
\hline
M: & 1713, Jan 08 &  7:17 & 0,467 &        &            \\
S: & 1713, Jan 22 & 11:14 & 0,626 &        &
     \raisebox{0.75\height}[0mm][0mm]{\cite{fru_kelley-milone}} \\
\hline
M: & 1659, Feb 09 &  6:21 & 0,294 &        &            \\
S: & 1659, Feb 23 & 12:23 & 0,748 & \raisebox{0.75\height}[0mm][0mm]{(long?)} & \\
\hline
M: & 1602, Dec 31 &  0:11 & 0,375 &        & \glqq last watch\grqq\ incorrect \\
S: & 1601, Jan 14 & 15:50 & 0,914 & \raisebox{0.75\height}[0mm][0mm]{(middle?)} &
         central + annular in Babylon \\
\hline
M: & 1547, Feb 01 &  1:56 & 0,264 &        &            \\
S: & 1547, Feb 15 & 13:53 & 0,809 &
     \raisebox{0.75\height}[0mm][0mm]{(short?)} &
     \raisebox{0.75\height}[0mm][0mm]{\cite{banjevic_2006}}\\
\hline
M: & 1416, Jan 23 &  3:56 & 1,196 &        &            \\
S: & 1416, Feb 07 & 13:25 & 0,123 &        &
     \raisebox{0.75\height}[0mm][0mm]{too late?} \\
\hline
M: & 1362, Feb 25 &  6:47 & 1,023 &        &            \\
S: & 1362, Mrc 12 & 10:24 & 0,587 &        &
     \raisebox{0.75\height}[0mm][0mm]{\cite{fru_mitchell} --- historically too late} \\
\hline
\hline
M: & 2161, Feb 09 &  3:44 & 1,278 &         &        \\
S: & 2161, Feb 25 & 17:17 & (0,4?)&
     \raisebox{0.75\height}[0mm][0mm]{long} &
     \raisebox{0.75\height}[0mm][0mm]{\cite{fru_huber_1999}: Akkad?} \\
\hline
M: & 2049, Feb 02 &  0:06 & 1,213 &         & \glqq last watch\grqq\ incorrect \\
S: & 2049, Feb 16 &  8:10 & (0,7?)&
     \raisebox{0.75\height}[0mm][0mm]{(short?)} &
     (\cite{fru_kelley-milone}: alternatively 2041?) \\ 
\end{tabular}
\end{table*}

In the generous time span from 1800 to 1300 BCE there are only
seven eclipse pairs, M + S, stored in the month of Shabattu.
They are listed in Table \ref{tab:babylonpaare}.
Penumbral eclipses are excluded for their invisibility.

When comparing the seven years with those of Table
\ref{tab:chronologies} for the sack of Babylon, the years 1753 and
1713 BCE appear historically too early.
They will only be feasible, if even longer chronologies are chosen
than the proposed ones.
At the beginning of the 20th century, when the EAE texts were
deciphered correctly, scholars were struggling with five different
ways of reckoning, at least.
They implied a fixing point for the capture of Babylon between
1977 and 1750 BCE \cite{fotheringham_1928}.
Today those models are outdated and not used anymore, since
they do not run conform to crosslinks with Assyria or Egypt.

At the other tail of the options, the pair of 1416 BCE appears
too late.
Wayne Mitchell prefers even the later solution of 1362 BCE
\cite{fru_mitchell}.
His calculation is said to provide excellent agreement from the
astronomical point of view, but the synchronisation with
archaeological evidence collapses completely.
Neither the king lists nor dendrochronology nor the connections to
Egypt harmonise with it.
For example, the Hittite conqueror of Babylon, Mur\v{s}ili I,
would be contemporary to the Egyptian pharaoh Akhenaten.
This is absolutely incompatible with historical ramifications.
The author of this study draws upon a theory of three solar
eclipses, in the 14th century, being responsible for Akhenatens's
admiration of the sun such that his lifetime sets a lower limit to
the choices for the Babylonian era \cite{khalisi-egypt}.
Mitchell, however, untightens the problem ostensibly by introducing
a ``super-short'' chronology with all data extremely compressed.
Many consider that too fancy.

The results for 1659 and 1602 BCE in Table \ref{tab:babylonpaare}
are difficult to justify, for the double eclipse occurs 8 or 7
years before its respective fix point.
Such large a discrepancy between the date and the supposed sack
of Babylon can hardly be explained.
The former interval harboured 11 lunar eclipses (8 partial and 3
total), though not in Shabattu, but they weaken the relationship
with the prophecy given on the tablet.
The astrologer would rather relate the ``fulfilment'' of the omen
to a closer eclipse and keep that for the record rather than an
event in the distance of eight years.

We are left with the pair of 1547 BCE, however, it does not cover
the scheme of the prevalent chronologies.
Boris Banjevic outlined an ``Upper Short Chronology'' such that
the system becomes more complicated \cite{banjevic_2006}.
He states that this solution would be consistent with the Venus
Tablet of Ammi-saduqa, but calculations by both Wayne Mitchell
and Peter Huber showed, a few years before, that the contrary is
true \cite{fru_mitchell, fru_huber_2000}.
The heliacal sightings of Venus would not coincide with the lunar
months, if Banjevic's year was chosen as the year of downfall.
Furthermore, Banjevic believes that the solar eclipse was merely
forecasted on a short timescale subsequent to the lunar eclipse.
This would mean that the solar eclipse does not need to be an
immediate one.
If it was not promptly observed, then it may have occurred in a
later month at a memorable space of time from the lunar eclipse.
That subverts the 14-day-distance in Shabattu (further
considerations below).

\begin{figure}[t]
\includegraphics[width=\linewidth]{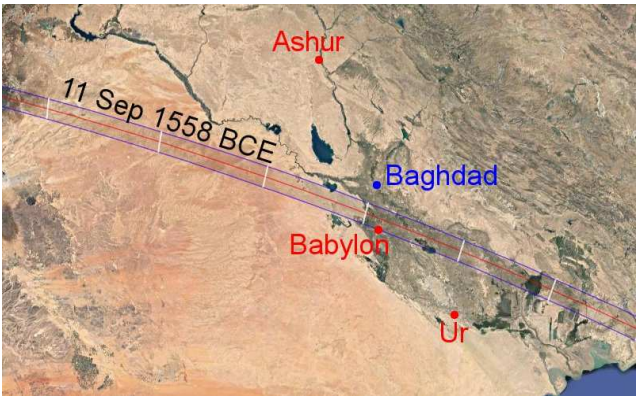}
\caption{Total solar eclipse of 11 September 1558 BCE after Espenak
    \cite{espenak}.}
\label{fig:eclipse1558}
\end{figure}

The archaeoastronomer G\"oran Henriksson disregards the omen on
the double eclipse and refers to the allusion of a single solar
eclipse on Tablet EAE \#24 \cite{henriksson_2005}.
He links it to 11 September 1558 BCE
(Fig.\ \ref{fig:eclipse1558}).
The totality could have passed over Babylon within the scope of the
shift due to $\Delta T$ (see Sec.\ \ref{ch:verschiebung}).
The central zone was as wide as 25 km, and the darkening occurred
at 11 a.m.\ local time.
Henriksson bases his arguments on the accompanying information on
that tablet describing the plundering raid of Babylon as well as
a devastating conflagration in a quarter of the city.
The dating rests upon geostratigraphy at excavation when the tablet
was discovered.
Using various additional information he builds up a bridge to a
new chronology similar to the short one.

\section{Inverted order of eclipses}

The two eclipses in Shabattu do not provide a satisfactory answer.
A way out could be perhaps an inverted sequence of the eclipse
types:
SM instead of MS.
For this option five candidates can be identified for the same
time span, see Table \ref{tab:inverted}.

\begin{table*}[t]
\caption{Inverted sequence of eclipse pairs in Shabattu visible in
    Babylon between 1800 and 1300 BCE.}
\label{tab:inverted}
%
\centering
\begin{tabular}{clrl|c|l}
\hline
\rowcolor{grey20}
\multicolumn{4}{c|}{\cellcolor{grey20}{Eclipse pairs SM:}}& Chrono- &   \\
\rowcolor{grey20}
   & Date         &Time [LT]& Mag. & logy &
                              \raisebox{0.75\height}[0mm][0mm]{Remarks} \\
\hline
S: & 1704, Jan 12 &  9:05 & 0,944 &        &            \\ 
M: & 1704, Jan 28 & 19:22 & 0,500 &        & at moonrise \\
\hline
S: & 1650, Feb 14 & 11:02 & 0,977 &        &            \\ 
M: & 1650, Mrc 02 & 17:49 & 0,730 & \raisebox{0.75\height}[0mm][0mm]{(long)} &
                                    at moonrise \\
\hline
S: & 1389, Feb 09 & 17:33 & 0,945 &        & central + annular at sunset \\ 
M: & 1389, Feb 24 & 23:01 & 0,224 &        & \glqq last watch\grqq\ incorrect \\
\hline
S: & 1351, Dec 30 &  9:53 & 0,251 &        &            \\ 
M: & 1350, Jan 15 & 16:05 & 1,268 &        & at moonrise \\
\hline
S: & 1335, Mrc 13 & 14:24 & 0,763 &        &            \\
M: & 1335, Mrc 28 & 22:45 & 0,523 &        & \glqq last watch\grqq\ incorrect \\
\end{tabular}
\end{table*}

\begin{figure}[t]
\centering
\includegraphics[width=\linewidth]{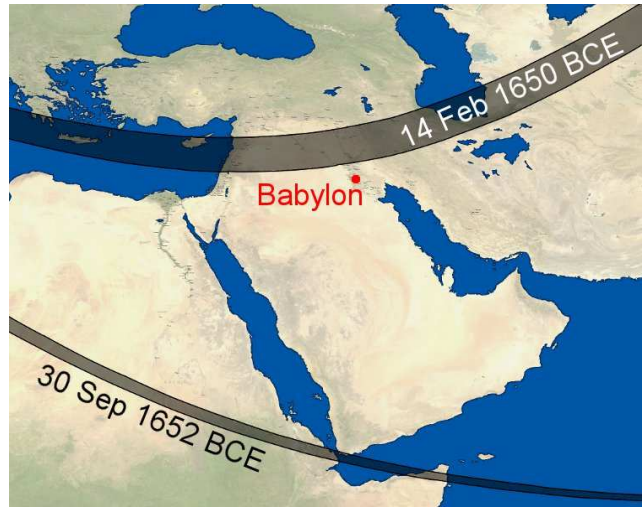}
\caption{Two solar eclipses accountable for the long chronology on
    the ``inverted'' sequence having the solar eclipse before the
    lunar.}
\label{fig:sackofbabylon}
\end{figure}

As above, the cases of the 14th century BCE, that might approach the
super-short chronology, must be rejected as historically too late.
At first sight, the event of 1650 BCE seems to be the sole feasible
candidate to meet the long chronology.
However, the interpretation comprises a good number of flaws.
On 14 February 1650 BCE, the observer would have seen a solar
obscuration with magnitude of 0.977 in Babylon.
The southern border of the totality zone passed by at a distance
of 120 km (Fig.\ \ref{fig:sackofbabylon}).
Two weeks later, on 2 March, a partial eclipse of the moon followed
with a magnitude of 0.730 happening in the evening time at
moonrise.
The time of day contradicts the text, as ``the last watch'' is
specified.
A scribal error must be excluded, for the subsequent two lines
speak of a simultaneous observation of the sun and moon.
Such a threefold reinforcement cannot be ignored.
The information on time receives high relevance, and the eclipsed
moon must have set at dawn, in particular, if the word
\textit{shurinnu} shall make sense.

But there is more to it.
These two eclipses in spring of 1650 BCE happened one year after
the presumed downfall of Babylon.
This impairs the astrological role of a ``prediction'', so the
omen itself becomes obsolete.

Even more important, the alteration of the order, SM, implies
that the eclipses occurred in two different Babylonian months.
Either the solar eclipse belongs to the previous month, or
Shabattu is grossly wrong for the lunar eclipse.
If Shabattu is erroneous, the search becomes hopeless.

The last resort would be a split of the eclipses into two different
windows of time being six months apart, at least, with only the
lunar eclipse belonging to Shabattu.
Such a separation will overstrain the arguments.
The number of possibilities increases rapidly, especially, when
picking up lesser obscurations.
We give an example for the dilemma that gets close to the date of
1651 BCE (not included in the table):
the total eclipse of the moon on 23 March 1652 BCE was rather small
(mag = 0.238), and the maximum obscuration was reached at 9 p.m.\
local time --- contrary to the text.
The next window of opportunity opened half a year later:
on 30 September 1652 BCE.
A partial solar eclipse could probably have been visible in Babylon
(Fig.\ \ref{fig:sackofbabylon}).
Its magnitude of 0.530 at 4 p.m.\ local time was not conspicuous,
and it could have escaped attention, either.
Solar eclipses with mag $<0.7$ are not necessarily noticeable,
even at best weather conditions.
Exploiting the error tolerance for $\Delta T$ (see Sec.\
\ref{ch:verschiebung} below) does not fundamentally improve the
boundary conditions for the benefit of its visibility.

The circumstances of weather are an important unknown, anyway.
This is often omitted when reasoning about eclipses.
It is taken for granted that the event was spotted, as the maps of
today present a convenient course of the moon's shadow over the
surface of the earth.
But the actual sighting of an eclipse is not self-evident at all!


\section{Akkad in Focus}
\label{ch:akkad}

An expert on history may probably be puzzled about the end of
the apodosis of that omen.
It is said that Ur took victory over Babylon. ---
Is that correct?

%
As pointed out in the historical overview, Babylon was conquered
by the Hittite king Mur\v{s}ili I as it was already debilitated.
The takeover is badly transmitted, and the sequence of Babylonian
rulers proceeds with Kassite kings.
A ``dark period'' of unknown length between the conquest and the
subsequent monarchs is widely assumed.
In contrast to that, the third and final dynasty in the city of
Ur ended two centuries prior to the rise of Babylon.
Ur was not extinguished yet, when the Babylonian Empire expanded,
but it lost most of its political importance.
If power was to be transferred from one city state to another,
then vice versa:
Babylon is the younger empire.

The statistician Peter Huber put forward a brilliant idea
\cite{fru_huber_1999}:
the omen needs to be shifted to 2161 BCE, and the long chronology
applied to Ur.
Under this perspective, he suspects that a mistake crept in
as ``Babylon'' will be a later insertion for ``Akkad''
(see lower part of Table \ref{tab:babylonpaare}).

Historically this is conceivable, indeed.
The Akkadian Empire of Sargon waned to a northern part of
Mesopotamia and was marked by social unrests.
On missing order the rulership passed over to a Gutian king.
The Gutians adopted all titles of Akkad and accommodated themselves
to the Mesopotamian society in many ways.
History appreciates them negatively, for their kings were obviously
unfamiliar with the urban economics such that supplies fell short.
That caused uproars among the population.
Civil wars were raging under all Gutian kings.
According to the Sumerian King List four kings changed within a
three year's time.
In the heartland, Gutium, their reign lasted 75 years or, possibly,
up to 100 till the king of Uruk, Utuhegal, drove them off from
several cities in Mesopotamia.
``Gutium'' itself is not identified, either.
It simply denotes a region, not a city, vaguely mapped to the
northeast of Akkad.
The last Gutian king was Tirigan, and he is said to have been in
power for 40 days
before being defeated and the kingship was taken to Uruk.

\begin{table*}[t!]
\caption{Suggestion for a timeline of the kings of Akkad, Uruk, and Ur.
    The eclipse omen EAE 20-XI is used as an anchor point for 2161 BCE.}
\label{tab:kingscheme}
\setlength{\arrayrulewidth}{0.8pt}
\centering
\begin{tabular}{r|c|c|c|c|l}
\hhline{*{6}{-}}        
\rowcolor{grey20}
Year [BCE]& ``Gutium'' &  Akkad    &     Uruk     &    Ur     & EAE-Omen \\
\hhline{*{6}{-}}        
       & \guti    & \multicolumn{2}{c|}{\cellcolor{yellow}} &   &  \\
       & \guti    & \multicolumn{1}{c}{\cellcolor{yellow}} & {\cellcolor{yellow}\dots\ (?)} &  & \\
         \hhline{~|>{\arrayrulecolor{braun}}-|>{\arrayrulecolor{yellow}}-|>{\arrayrulecolor{black}}-|~|~} 
 ????  & \guti    & \cellcolor{yellow}Sharkalisharri & \urukiv  &  & EAE 21-VIII? \\
         \hhline{~|>{\arrayrulecolor{braun}}-|>{\arrayrulecolor{black}}-|%
                  >{\arrayrulecolor{blaua}}-|>{\arrayrulecolor{black}}~|~} 
       & \guti \raisebox{1mm}[0mm][0mm]{\fbox{Gutian period}} &   &
                                  \urukiv \raisebox{1mm}[0mm][0mm]{\fbox{Uruk-IV-Dynasty}} &   & \\
       & \guti    &             & \urukiv \dots &          & \\
         \hhline{~|>{\arrayrulecolor{braun}}-|>{\arrayrulecolor{black}}~|%
                  >{\arrayrulecolor{blaua}}-|>{\arrayrulecolor{black}}~|~} 
         \cdashline{5-5}[0.5pt/3pt]  
       & \guti {\it (total of 21 kings} & \dots\ (?) & \urukiv Ur-Gigir (6 years) &  & \\
         \hhline{~|>{\arrayrulecolor{braun}}-|>{\arrayrulecolor{black}}~|%
                  >{\arrayrulecolor{blaua}}-|>{\arrayrulecolor{black}}~|~} 
\dots  & \guti {\it in $\approx$86--96 years} & & \urukiv Kuda (4 years) & & \\
         \hhline{~|>{\arrayrulecolor{braun}}-|>{\arrayrulecolor{black}}~|-|~|~} 
 2171  & \guti {\it in the land of Guti)} &  & \urukiv Puzur-ili (5 years) &  & \\
 2170  & \guti    &             & \urukiv      &           & \\
 2169  & \guti \dots &          & \urukiv      & (Ka-du?)  & \\
 2168  & \guti Yarlaganda (7 years) & & \urukiv &          & \\
         \hhline{~|-|~|>{\arrayrulecolor{blaua}}-|>{\arrayrulecolor{black}}~|~} 
 2167  & \guti Si'um (7 years) & & \urukiv Puzur-ili &     & \\
         \hhline{~|>{\arrayrulecolor{braun}}-|>{\arrayrulecolor{black}}~|-|~|~} 
 2166  & \guti    & {\it (4 kings in} & \urukiv Ur-Utu (6 years) &  & \\
 2165  & \guti    & {\it 3 years?)} & \urukiv  &           & \\
 2164  & \guti    &             & \urukiv      &           & \\
 2163  & \guti    &   Igigi     & \urukiv      & (E-lili?) & \\
 2162  & \guti    & Nanum, Imi  & \urukiv      &           & \\
{\bf 2161} & \guti Si'um | Tirigan & \guti Elulu &
                                  \urukiv Ur-Utu | \color{braun}{Tirigan} & & {\bf EAE 20-XI}\\
         \hhline{~|>{\arrayrulecolor{braun}}-|-|-|>{\arrayrulecolor{black}}-|~} 
 2160  &          & \urnammu Dudu (21 yrs?) &
                                  \urukv Utuhegal & \urnammu Ur-Nammu & EAE 21-IV \\
    \hhline{~|~|>{\arrayrulecolor{hellblau}}-|>{\arrayrulecolor{blaub}}-|>{\arrayrulecolor{hellblau}}-|>{\arrayrulecolor{black}}~}  
 2159  &          & \urnammu {\dots } & \urukv & \urnammu (as governor)& \\
 2158  &          & \urnammu Shu-Durul (15 yrs?) & \urukv  & \urnammu  & \\
 2157  &          & \urnammu {\dots } & \urukv & \urnammu  & \\
 2156  &          &             & \urukv \raisebox{1mm}[0mm][0mm]{\fbox{Uruk-V-Dynasty}}
                                               & \urnammu  & \\
 2155  &          &             & \urukv       & \urnammu  & \\
 2154  &          &             & \urukv       & \urnammu  & \\
 2153  &          &             & \urukv       & \urnammu  & \\
 2152  &          &             & \urukv       & \urnammu  & \\
 2151  &          &             & \urukv       & \urnammu  & \\
 2150  &          &             & \urukv Utuhegal & \urnammu & EAE 20-IV? \\
         \hhline{~|~|~|-|~|~} 
 2149  &          &             &\multicolumn{2}{c|}{\urdrei Ur-Nammu (as king)} & \\
 2148  &          & \cellcolor{hellblau}{\dots } &\multicolumn{2}{c|}{\urdrei} & \\
 2147  &          & \cellcolor{hellblau}{End of Akkad?} &\multicolumn{2}{c|}{\urdrei} & EAE 21-I? \\
    \hhline{~|~|-|>{\arrayrulecolor{tuerkis}}-|>{\arrayrulecolor{tuerkis}}-|>{\arrayrulecolor{black}}~} 
 2146  &          &             &\multicolumn{2}{r|}{\urdrei \fbox{Ur-III-Dynasty}} & \\
 2145  &          &             &\multicolumn{2}{c|}{\urdrei} & \\
 2144  &          &             &\multicolumn{2}{c|}{\urdrei} & \\
 2143  &          &             &\multicolumn{2}{c|}{\urdrei} & \\
 2142  &          &             &\multicolumn{2}{c|}{\urdrei Ur-Nammu (10 + 8 years?)}& \\
         \hhline{~|~|~|-|-|~} 
 2141  &          &             &\multicolumn{2}{c|}{\urdrei \v{S}ulgi (46 years)} & \\
\dots  &          &             &\multicolumn{2}{c|}{\urdrei} & \\
       &          &             &\multicolumn{2}{c|}{\urdrei} & \\
\end{tabular}
\end{table*}

Based on this concept, that eclipse omen would deal with the king
Utuhegal (Table \ref{tab:kingscheme}).
He would have come to power in Uruk in 2161 BCE where he founded
the fifth dynasty whose sole member he was.
He installed new governors in the cities under his control.
For example, his son-in-law was Ur-Nammu and administered Ur for
the next ten years.
In the same year or the year after, Utuhegal might have led the
revolt against the Gutians and, finally, attained hegemony over
Akkad.

The circumstances of the upheaval in Akkad are scanty,
unfortunately.
They are open for speculations.
According to some sources the last regular Akkadian king was
Sharkali\-sharri, but he lost a number of cities in the area of
Sumer on account of a drought,
maybe Akkad itself.
After him the king list mentions four potentates who were vying to
be king.
There is no evidence from this short interval, and we do not know
anything about the contenders to the throne except their names.
Thoughts were raised whether they belonged to the Gutians making
inroads into Akkad.
Ilulu, the last one of them, was almost certainly a Gutian,
before Dudu seized power over the city after those three years of
confusion \cite{macqueen_1964}.

King Dudu is described as Sharkalisharri's successor, however,
without a family kinship to him.
After 21 years in office Dudu was replaced by his son Shu-Durul.
The latter kept himself on the throne for 15 years, and he is
deemed to be the final ruler of Akkad.
According to classical teaching, it was only after Shu-Durul that
the Gutians took over control in Akkad.

The chaotic period after Sharkalisharri is doubtful in both length
and persons involved.
An intriguing question concerns the affiliation of Dudu and
Shu-Durul.
Were they Akkadians?
Were they really kings
or rather governors, like Ur-Nammu, appointed by Utuhegal? ---
Dudu's rule was limited to a little more than the capital itself.
There are no eponym years known from his time, and, in general,
it seems unlikely that he could have reigned as long as 21 years.
It remains unclear what person was operating in which city in
whose charge.

Form another source we learn that Utuhegal, who won victory over
the Gutians, perished tragically in the 7th year of his regency,
instead of the tenth year, while visiting a dike
\cite{fru_huber_2000}.
After his death, Ur-Nammu, the son-in-law, seized the opportunity
to ascend the throne of Ur, which he had controlled before as
governor.
At the same time he might have taken command of Uruk itself and,
possibly, the orphaned city of Akkad that would have been part of
Utuhegal's realm.
It is quite certain that Ur-Nammu founded the third dynasty of Ur,
and reigned there as king for a minimum of eight years.
It may be conjectured that his regency was counted from his
governorship instead of accession as king, resulting 18 years in
total.
The commencement would be put in the year of the double eclipse
of 2161 BCE.

Assigning the omen \#20-XI to Akkad, as suggested by Huber, some
further lunar eclipses on the same tablet could yield potential
solutions.
Other apodoses could be contemplated a transition in the
Ur-III-Dynasty.
A number of kings would have died shortly after an eclipse, and
each of them was substituted by his son, see \cite{fru_huber_1999}.
This could apply to the transition from Dudu to Shu-Durul, too,
if we conjecture, somewhat arbitrarily, a much shorter time for
them, tentatively 15 years for their combined reign and allot the
omen \#21-I to the demise (see Appendix). 
That text characterises a smooth abandoning of Akkad without a
warfare.

The king scheme in Table \ref{tab:kingscheme} is destined to be
improved in the future with the omens in the rightmost column
guiding the way.
Taking 2161 BCE as the anchor, it supports the long chronology
still.
%
From Table \ref{tab:babylonpaare} can be inquired that the short
chronology probably offers another solution for the double eclipse
in 2049 BCE, but the historical details turn out vague, so we will
abstain from a discussion.
According to Kelley \& Milone \cite{fru_kelley-milone}, the year
2041 BCE is supposed to supply an alternative,
however, we did not find a suitable pair of eclipses.
Glancing at the ultra-short chronology, there was a lunar and a
solar eclipse on 28 April and 12 May 1905 BCE, respectively, but
the dates are much too late in the season to meet the month of
Shabattu.



\section{Alternative timeline for Akkad}

Different from Table \ref{tab:kingscheme}, there are other ways of
scheduling the kings in Akkad.
The classic draft embodies a strict adoption of the regencies of
those four short-time kings.
If the 3 years were true, the column for Akkad would have to be
squeezed and moved upward in time.
All columns are slidable in vertical direction while preserving the
succession of kings.

Currently, the end of Sharkalisharri (at the ``????'') is estimated
at about 2210 BCE on the long chronology and 2190 BCE on the middle
chronology \cite{macqueen_1964}.
However, the time gap between Sharkalisharri and Ur-Nammu could be
very small, as the orientalist Claus Wilcke asserts,
probably one generation only, see quotation by Huber
\cite[p64]{fru_huber_1999}.
In any case, historians would have to view the Gutian period in a
new light.
Eclipse years are able to jump upwards or downwards at the Exeligmos
cycle (54 years, see \cite{finsternisbuch}) with only minor
difficulties.

Still, Utuhegal's victory over the Gutians is free to be combined
with a completely different omen.
For instance, the omen on tablet EAE \#21-IV deals with a lunar
eclipse in the month IV (about June--July).
The apodosis reads \cite{fru_huber_1999}:
\begin{quote}
If an eclipse occurs on the 14th of Dumuzu (IV), \dots\
it begins in the evening watch and clears in the middle watch \dots\
The prediction is given for the king of Guti:
The downfall of Guti in battle.
The land will be totally laid waste.
\end{quote}

Here, the Gutians are directly addressed, but we do not know to
whom it applies.
What king?
Which land?
Where is ``Guti''?
Following Peter Huber, the eclipse of 24 July 2160 BCE would match
this omen of EAE \#21-IV.
The solution would not suspend our former double eclipse of EAE
\#20-XI, but it would weaken its significance.

Another speculative result, wherein both omens \#20-XI and \#21-IV
could find a historical counterpart, would be that Utuhegal first
terminated the Gutian rule in Uruk or ``Gutium'', and then
combated Akkad one year later, i.e.\ both kings Elulu and Dudu
would slip one year lower in Table \ref{tab:kingscheme}.
This reasoning appears quite tenuous in light of lacking evidence.
As long as there are no names given in the apodoses, many omens
are exchangeable with respect to their interpretation.
The portents seem to be a collection of events spread over
700 years \cite{cornelius_1969}.



\section{Irregular Rotation of the Earth}
\label{ch:verschiebung}

The conjecture on a modified text on EAE \#20-XI does not offer
the only answer to the problem.
In his investigation Peter Huber implements the lunar eclipse only
and omits the eclipse of the sun on 28th of Shabattu.
The associated solar eclipse, that would correspond to 25 February
2161 BCE, was most likely not visible in Mesopotamia.
It pertained to a partial coverage far away to the north of Europe.
The axis of the shadow of the moon passed by the North Pole off
the earth ($\gamma$ = 1,0965).
The clock-time error, called $\Delta T$, is estimated to
50,397$\pm$4,430 seconds, whereat the error of tolerance reveals
the crucial value.

The clock error $\Delta T$ is an indicator for the deceleration
of the rotation of Earth caused primarily by tidal friction.
The quantity denotes the difference between the strictly uniform
timescale (TT), measured with atomic clocks, and the constantly
lengthened timescale for the day (UT), which is used for our civil
information on time:
\begin{eqnarray*}
\Delta T &=& \text{TT} - \text{UT} \\
         &=& -20 + c \cdot t^2 \;\; \text{[s]},
\end{eqnarray*}
where the constant $c \approx 32$ s/(cy)$^2$ and $t$ is given in
centuries (cy) before 1800.
For the detailed geoscientific background on the effect of
deceleration see \cite{finsternisbuch} or respective literature.

The value of $\Delta T$ is known quite reliably from Antiquity
($\approx$700 BCE) till now, but for epochs further back in time
it is extrapolated.
The extrapolation deploying the formula above considers the
\emph{regular} and systematic slow-down that accounts for about
2 milliseconds per day.
But it does not include the minute random fluctuations arising
from the unanticipated behaviour.
For example, a climatic melt of the polar ice would increase the
ocean level, or a strong earthquake may lead to a displacement of
continents.
Both effects alter the moment of inertia of the earth, and its
rotation is accelerated (positive or negative with respect to the
average) in an unpredictable way.
These changes are tiny, but they do accumulate over centuries to
a conspicuous error.
Thus, when the astronomical conditions for an eclipse are fulfilled,
the earth presents a different surface to the celestial actors.
A precise backcalculation in terms of local time cannot be
guaranteed;
the lunar eclipse will be shifted to a different time zone.

The geographical shift takes effect on solar eclipses even more.
In our example of 25 February 2161 BCE, the uncertainty amounts to
4430 s = 74 min, meaning that the visibility track could pass a
certain geographical longitude within that error of time earlier
or later.
The eclipse itself lasts for about 90 minutes, and statements on
observational conditions become impossible:
perhaps the partial eclipse was seen at a small magnitude in
Mesopotamia, perhaps it began after the earth has turned away to
the night side.
In the latter case it would have taken place below the horizon and
must be excluded from visibility.

The \emph{average} $\Delta T$ for 25 February 2161 BCE yields a
maximum phase near Iceland (mag $\approx$0.4).
In order to generate the largest obscuration of the sun's disk for
Mesopotamia, a higher $\Delta T_{\rm opt}$ of $\approx$65,000 s is
required.
This targeted value is going to exceed the error bar by the factor
of three and, thus, hard to justify.

\begin{figure}[t]
\includegraphics[width=\linewidth]{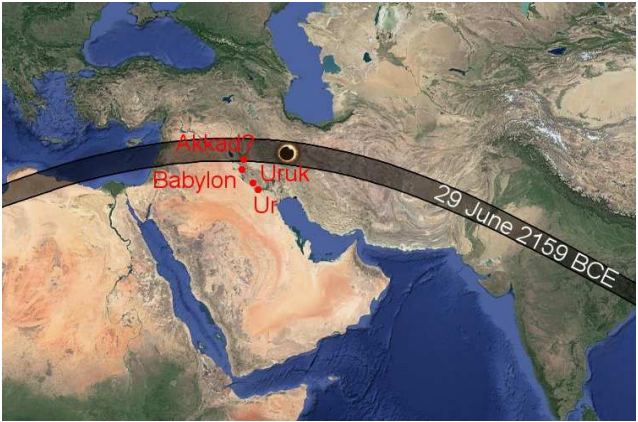}
\caption{Course of the solar eclipse of 29 June 2159 BCE.
    The sun symbol marks the spot of maximum totality.}
\label{fig:akkad-sofi}
\end{figure}

Now, let us put the separation of the two eclipses in Shabattu to
the test and look whether the solar eclipse would fit a later month.
An impressive event can be discovered two and a half years after
the lunar (Fig.\ \ref{fig:akkad-sofi}).
Using the average $\Delta T$, the sun suffered an eclipse on 29
June 2159 BCE at noon.
The totality was even exceptionally long, as darkness lasted for
about 6 minutes.
The assumed Akkad ($\approx$ Baghdad?) resided at the edge of the
central zone, and there was a magnitude of $\approx$0.93, at least,
in Uruk and Ur.

However, there is a disturbance in Huber's Akkad hypothesis:
The word ``Babylon'' recurs four times in the text.
This cannot be attributed to an accidental scribal error.
The suspicion of a deliberate amendment makes the state of affairs
more complicated:
Who was to do so and for what reason?
Also, the space of two and a half years of idle time between the
lunar and solar eclipse remains unattractive.

On the other side, the systematisation of astrological omens tends
to support Huber's Akkad hypothesis.
In the course of many generations the divinations would have
received a gradual evolution.
An overall assumption is that any kind of celestial documentation
began with eclipses once, and a first catalogue of various
portents was put to record.
Some of them would have ``come true'', e.g.\ the death of a regent
or a social misery, so the incidents strengthened the belief in
the doom prophesied by celestial signs.
The passage on the double eclipse of EAE \#20 would have mutated
to a What-if-speculation.
Hence, it does not necessarily deal with a historical fact.


\section{Conclusions}

We analysed the omen on EAE Tablet \#20 in Shabattu in regard to
three possible interpretations:
literal comprehension, inverted order of eclipses, and a transfer
of the historical scene in the apodosis.
Each variant shows serious shortcomings.
But no matter how to turn it, the long chronology is better off
than any other historical scale.
The least obstacles emerge, when the lunar eclipse is assigned to
the end of the Gutian rule in Akkad, as proposed by Huber, instead
of the original wording ``Babylon''.

That cuneiform tablet provides just \emph{one} attempt to shed
light on the historical stage.
There are manifold pieces from other disciplines that would cast
deviant solutions.
The middle chronology is the one that bears the worst concordance
with astronomy,
but it is so deeply rooted in the minds of historians that it will
almost certainly continue to be used, though there is no supporting
evidence for it.
It serves rather as a compromise to suit as much authors as
possible.
This comes along ``democratic'', but it is not scientific.

Actually, astronomy is the field that provides the hardest facts
to chronology.
The radiocarbon method and dendrochronology could also provide
reliable time slots, but both are subject to uncertainties of
their own.
It concerns the material available:
the origin of the sample is not known in many cases, or its
attribution to the culture fails, or it might be contaminated.
In addition, a much larger error in measurement (confidence
interval) has to be accepted.
Texts are sometimes subject to doubtful retrieval.
And the interpretation itself often arises from a rather modern
understanding than from knowledge of the ancients.
At this point a combination of astronomical, archaeological, as
well as historical evidence will be of high value.


\section*{Acknowledgements}

This paper is an excerpt from the Habilitation submitted to the
University of Heidelberg, Germany \cite{finsternisbuch}.
The unpaid research was accomplished under direful
circumstances.
The relevant section is revised and now published via
{\tt arXiv} without assessment.


\section*{Appendix}
\label{ch:app}

Further texts (abbreviated) from the EAE tablets assigned in
Table \ref{tab:kingscheme}.
Dates proposed for lunar eclipses after \cite{fru_huber_1999}.

\noindent
\paragraph{EAE 21-VIII:}
``If an eclipse occurs on the 14th day of Arahsamna (month VIII) \dots\
The prediction is given for the king of the world:
Either the king will die, or a large army will fall, or a large
army will revolt.''\\ [0.5\baselineskip]
\textit{Possible context:}
The decline of the Akkadian Empire set in with Sharkalisharri who
lived about 150 years after Sargon.
Control was lost over large parts of the country, while neighbours
increased political and military pressure. ---
Alternatively, that omen can be made conform with Utuhegal's death,
see below EAE 20-IV.\\ [0.5\baselineskip]
\textit{Lunar Eclipses:}
8 Jan 2150 $\;$ or $\;$ 10 Jan 2085 $\;$ or $\;$ 7 Dec 2186 BCE

\vspace{\baselineskip}
\noindent
\paragraph{EAE 20-IV:}
``If an eclipse occurs on the 14th of Dumuzu (IV) \dots\
The king who ruled will die. \dots\
The prediction is given for Ur. \dots\
The grandson, descendant of the king, will seize the throne. \dots\
The king together with his clan will be killed.''\\ [0.5\baselineskip]
\textit{Possible context:}
A skip from a king to his grandson is unknown for the entire
dynasties of Akkad and Ur-III.
All kings were sons of their predecessors.
At best, a transition to a brother did take place.
Arguable is also the death of Utuhegal, as he was followed by
his son-in-law Ur-Nammu.\\ [0.5\baselineskip]
\textit{Lunar eclipses:}
4 Jul 2150 $\;$ or $\;$ 25 Jul 2095 BCE

\vspace{\baselineskip}
\noindent
\paragraph{EAE 21-I:}
``If an eclipse occurs on the 14th of Nisannu (I) \dots\
The prediction is given for the king of Akkad.
The king of Akkad will die.
If the eclipse does not affect the king:
There will be destruction and famine.
The people will send their children out to the market (to be sold).
The great country will go to the small country for
food.''\\ [0.5\baselineskip]
\textit{Possible context:}
This could be understood as the end of the Akkad dynasty.
The two final kings were Dudu and his son Shu-Durul.
However, their regencies do not fill well to the
timeline, see Sec.\ \ref{ch:akkad}.\\ [0.5\baselineskip]
\textit{Lunar eclipses:}
3 May 2147 $\;$ or $\;$ 31 Mrc 2201 BCE \\




\end{document}